\documentclass[a4paper, 10pt, conference]{ieeeconf}      

\IEEEoverridecommandlockouts                              

\usepackage{graphicx} 
\usepackage{xcolor}
\usepackage{subcaption}
\usepackage{url}

\usepackage{fancyhdr} 
\usepackage{setspace} 

\title{\LARGE \bf
Analyzing Data Efficiency and Performance of Machine Learning Algorithms for Assessing Low Back Pain Physical Rehabilitation Exercises
}

\author{Aleksa Marusic$^{1}$, Louis Annabi$^{1}$, Sao Mai Nguyen$^{1, 2}$ and Adriana Tapus$^{1}$
\thanks{*This work was supported by ENSTA Paris}
\thanks{$^{1}$Autonomous Systems and Robotics Lab, Computer Science and System Engineering (U2IS), ENSTA Paris, Institut Polytechnique de Paris, 828 Blvd des Maréchaux, 91120 Palaiseau, France, {\tt\small name.surname@ensta-paris.fr}}%
\thanks{$^{2}$Dep. Informatique, IMT Atlantique, {\tt\small nguyensmai@gmail.com}}%
\thanks{979-8-3503-0704-7/23/\$31.00 ©2023 European Union}
}

\begin{document}

\maketitle
\thispagestyle{fancy}
\lhead{}
\chead{
\texttt{
\begin{spacing}{0.9}
\vspace{-20pt}
\scriptsize{
Marusic, A., Annabi, L., Nguyen, S. M., and Tapus, A. (2023). Analyzing Data Efficiency and Performance of Machine Learning Algorithms for Assessing Low Back Pain Physical Rehabilitation Exercises. European Conference on Mobile Robots.  \url{https://dx.doi.org/10.1109/ECMR59166.2023.10256318}
 }
 \end{spacing}
}
\vspace{20pt}}
\rhead{}
\cfoot{}

\begin{abstract}

Physical rehabilitation focuses on the improvement of body functions, usually after injury or surgery. Patients undergoing rehabilitation often need to perform exercises at home without the presence of a physiotherapist. Computer-aided assessment of physical rehabilitation can improve patients' performance and help in completing prescribed rehabilitation exercises. In this work, we focus on human motion analysis in the context of physical rehabilitation for Low Back Pain (LBP). As 2D and 3D human pose estimation from RGB images had made impressive improvements, we aim to compare the assessment of physical rehabilitation exercises using movement data acquired from RGB videos and human pose estimation from those. In this work, we provide an analysis of two types of algorithms on a Low Back Pain rehabilitation datasets. One is based on a Gaussian Mixture Model (GMM), with performance metrics based on the log-Likelihood values from GMM. Furthermore, with the recent development of Deep Learning and Graph Neural Networks, algorithms based on Spatio-Temporal Graph Convolutional Networks (STGCN) are taken as a novel approach. We compared the algorithms in terms of data efficiency and performance, with evaluation performed on two LBP rehabilitation datasets: KIMORE and Keraal. Our study confirms that Kinect, OpenPose, and BlazePose data yield similar evaluation scores, and shows that STGCN outperforms GMM in most configurations.

\end{abstract}

\section{INTRODUCTION}

Physical rehabilitation has a very important role in postoperative recovery and in the restoration of body functions \cite{Wu20}. Usually, during the rehabilitation process, patients performing exercises are monitored in a clinical setting by a medical professional, such as a physiotherapist. During a rehabilitation exercise session, patients’ behavior reflects their health status and is an important indicator of the treatment outcome. However, patients often have a limited number of supervised sessions, and they need to continue the rehabilitation process at home without any supervision. In these cases, a physiotherapist makes a rehabilitation plan consisting of several recommended exercises. Patients are typically responsible for performing their exercises regularly at home and periodically visiting the hospital for progress assessment. However, a lack of supervision and timely feedback from healthcare professionals can reduce patient's engagement during the rehabilitation process. Lower motivation and poor supervision can increase the chances of incorrect exercise performance, which can slow down the recovery process and increase the risk of re-injury \cite{liao2020review}.

Low back pain (LBP) is a major cause of disability worldwide, with more than 50\% of the global population experiencing LBP at some point in their lives \cite{Wu20}. This is especially concerning as LBP disproportionately affects elderly individuals, whose percentage in European societies is steadily increasing. As a result, medical staff are under significant strain to manage the growing number of patients suffering from LBP.

Automatic physical rehabilitation monitoring can significantly improve patients' progress during at-home rehabilitation. The goal of such a system is to recognize the activity being performed, the intensity with which it is performed, and its quality, thus helping monitor patients' progress. In general, human activity analysis is a very active research topic today and one of the most important and challenging areas in AI. It involves analyzing human body movements based on the motions of different body joints, skeletons, and muscles \cite{Sardari2023}. It also has applications in several domains such as sports sciences, action or gesture recognition \cite{Kulic2009DetectingCI, Aggarwal2011, Glowinski2011, Tapus2018}, and range-of-motion estimation \cite{Miron2021}. 

Developing an effective system for movement assessment highly depends on a few factors including motion sensors, precise movement data and its pre-processing, and evaluation techniques. In recent years, there were several studies that employed machine learning methods to classify individual repetitions into correct or incorrect classes of movements. Some of the first methods proposed for this task included distance function-based algorithms such as Dynamic Time Warping and Mahalanobis distance or probabilistic models such as hidden Markov models and Gaussian mixture models \cite{Su2014KinectenabledHR, Capecci2018}. The outputs in these approaches are discrete class values of 0 or 1 (i.e., incorrect or correct). 


Naturally, with recent developments in Neural Networks and Deep Learning (DL), there is a big interest in their application for modeling and analysis of human motions. There are already numerous papers on general Human Action Recognition (HAR) systems that utilize various DL frameworks ranging from Convolutional Networks and Long Short-Term Memory (LSTM) \cite{Gu2021} and encoder-decoder networks to even more novel architectures such as Spatio-Temporal Graphs \cite{yan2018spatial}, and Attention models \cite{Liu2022}.

Furthermore, a large number of datasets related to HAR fields are freely available for analysis. These datasets are extensively used for benchmarking algorithms for action recognition, gesture recognition, or pose estimation. However, in the medical domain, collecting large data sets of rehabilitation exercise data from patients faces multiple challenges such as impairment, unlabeled data, or privacy and safety concerns. Consequently, only a few public datasets for rehabilitation evaluation are currently available, and they are still quite smaller than more general ones, e.g. the ones used for benchmarking HAR algorithms. 

Our study focuses on evaluating the performance and data efficiency of two distinct algorithms used to assess the effectiveness of rehabilitation exercises. Specifically, we investigate the Gaussian Mixture Model and the Spatio-Temporal Graph Convolutional Network algorithm. Our evaluation is conducted on two datasets that contain rehabilitation exercises for Low Back Pain patients, namely Kimore \cite{capecci2019kimore} and Keraal dataset \cite{Nguyen2024IJCNN}.

The rest of the paper is structured as follows. Section \ref{sec:related_work} presents an overview of the different approaches for motion analysis and physical rehabilitation assessment. Section \ref{sec:dataset} describes the used datasets, while Section \ref{sec:method} details the implemented methods for rehabilitation assessment. The results are summarized in Section \ref{sec:results}. Finally, the conclusions and discussions are presented in Section \ref{sec:conclusion}.

\section{RELATED WORK}
\label{sec:related_work}

This section presents related work in deep learning for motion analysis in general and in the assessment of physical rehabilitation exercises.

\subsection{Deep learning for motion analysis}

Several deep learning approaches have been applied on skeleton data, in particular on the task of action recognition. Motion data can be seen as 3D tensors with one temporal dimension (the timeframe of the movement), one spatial dimension (the skeleton joint), and one feature dimension (the XYZ Euclidean position). Early deep learning approaches for motion processing focused on the temporal processing, using recurrent neural networks \cite{devanne2019recognition}, or 1D convolutional neural networks \cite{kim2017interpretable}. Other approaches explored the idea of representing motion as images, in order to exploit the performances of 2D and 3D convolutional neural network for image processing \cite{wang2016action, ke2017new, li2017joint, caetano2019skeleton, duan2022}.

With the recent development of graph neural networks, there is a new wave of algorithms that are able to properly take into account the skeleton structure using graph convolutions \cite{yan2018spatial, shi2019skeleton, liu2020disentangling}. These neural network layers perform an operation that can be seen as a message passing between adjacent joints in the skeleton graph, thus properly exploiting this prior knowledge. More recently, self-attention mechanisms have been added to allow graph convolutions to span across non-adjacent joints based on dynamically computed attention coefficients \cite{plizzari2021skeleton, mazzia2022action}.


\subsection{Assessment of physical rehabilitation exercises}

Movement assessment is typically accomplished by comparing a patient's performance of an exercise to the desired performance as specified by therapists. A sequence of body movements is provided as input for a machine or deep learning algorithm, which should assess that exercise with a quality score. This thus requires a more precise model of the movement than most gesture classification models.

At first, studies on exercise evaluation employed more traditional machine learning methods for classification, such as Adaboost classifier, K-Nearest Neighbors, Bayesian classifier, or ANNs \cite{Ilktan2014}. Others tried using distance function based models like \cite{Houmanfar2016MovementAO}. However, classifiers only provide correct or incorrect labels, not providing any additional information or score, while distance functions solve that problem but are not able to learn from the rehabilitation data.

Further, some of the research tried using probabilistic approaches, like Hidden Markov models \cite{Capecci2018} or Gaussian Mixture Model \cite{Nguyen2016IWCR}. Such models provide an assessment that is based on the likelihood that the given exercises are being drawn from a trained model. These models were able to solve previous problems, and stochastic character of human movements goes hand in hand with models nature, but they are not able to extract all the information from the data, such as joint or spatial connections among body parts.

Liao et al. \cite{Liao2020DLPRassessment} created a deep neural network model to generate quality scores of input movements. They proposed deep learning architecture for hierarchical spatio-temporal modeling combining GMMs, CNNs, and LSTM to provide a quality score. However, with recent development of Graph Neural Networks, it is possible to extract even more information from spatio-temporal features of the exercise. In \cite{deb2022graph} and \cite{Du2021AssessingPR}, Graph Convolutional Networks (GCN) are used to assess physical rehabilitation, obtaining state-of-the-art scores on commonly used KIMORE and UI-PRMD datasets. Last but not least, \cite{Yu2022} et al. used an ensemble of two GCN, one for position and for orientation features of the skeleton joints.

\section{DATASET}
\label{sec:dataset}

This section explains the type of data used and presents two datasets used for the evaluation of the algorithms.

\subsection{Skeleton data}

\begin{figure*}
     \centering
     \begin{subfigure}[b]{0.275\linewidth}
         \centering
         \includegraphics[width=\textwidth]{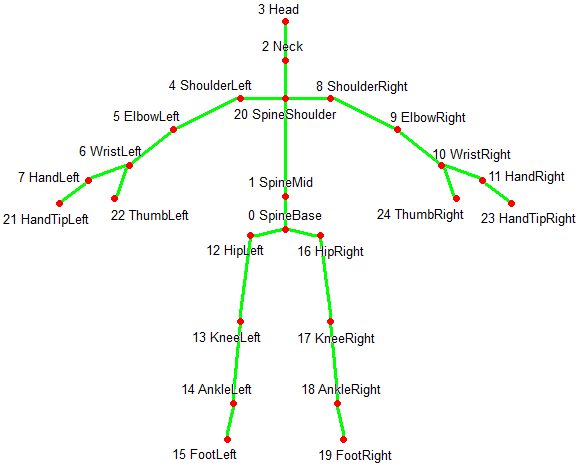}
         \caption{Kinect v2 - 3D positions and orientations of each joint of Kinect skeleton }
         \label{fig:kinect_skeleton}
     \end{subfigure}
     \hfill
     \begin{subfigure}[b]{0.325\linewidth}
         \centering
         \includegraphics[width=\textwidth]{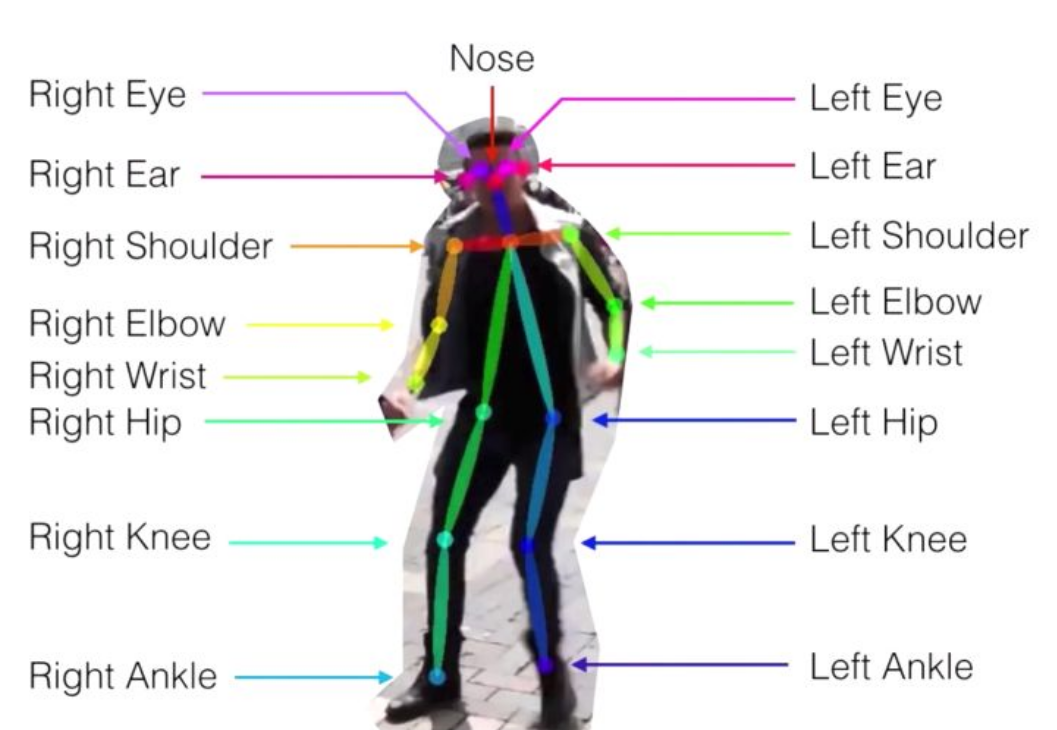}
         \caption{OpenPose - 2D positions of each joint of OpenPose skeleton}
         \label{fig:openpose_skeleton}
     \end{subfigure}
     \hfill
     \begin{subfigure}[b]{0.325\linewidth}
         \centering
         \includegraphics[width=\textwidth]{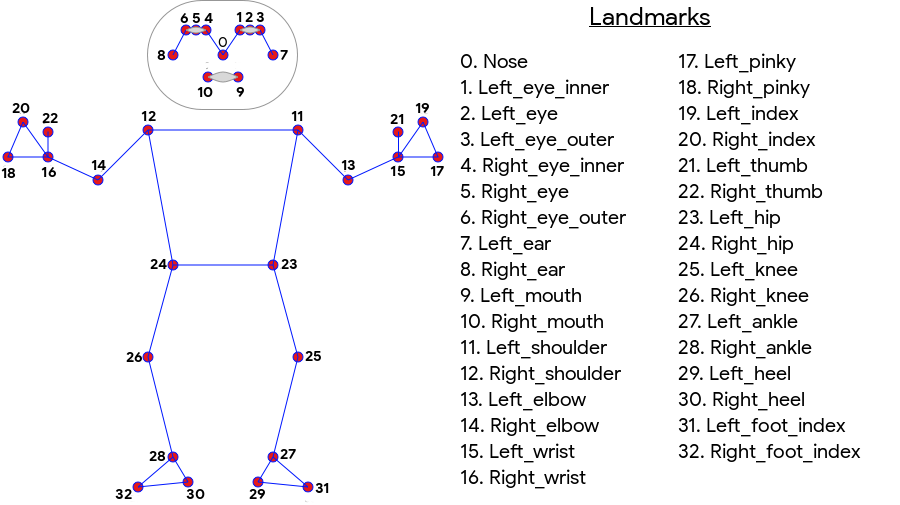}
         \caption{BlazePose - 3D positions of each joint of BlazePose skeleton}
         \label{fig:blazepose_skeleton}
     \end{subfigure}
  \vspace{-0.4em}
        \caption{Skeleton format for the three pose estimation algorithms used in the Keraal dataset. For the Kimore dataset, we only have Kinect data.}
        \label{fig:skeleton_data}
\end{figure*}

A Human Pose Skeleton represents the orientation of a person in a graphical format. 
Depth cameras, like the Microsoft Kinect, can provide position and orientation of skeleton joints. They had become very popular due to their price and ease of use over optical motion tracking systems, which place a set of markers on the body. 
More recently, a standard vision camera can be used with deep learning techniques that estimate skeleton joints positions from plain RGB images. In this work, we will consider the algorithms OpenPose and BlazePose \cite{Cao2019ITPAMI,Bazarevsky2020C}.  
Figure \ref{fig:skeleton_data} displays the joints of Kinect \footnote{\url{https://www.sealeftstudios.com/blog/blog20160708.php}}, OpenPose \footnote{\url{https://maelfabien.github.io/tutorials/open-pose/}} and BlazePose \footnote{\url{https://ai.googleblog.com/2020/08/on-device-real-time-body-pose-tracking.html}} skeletons.


\subsection{Keraal dataset}
The Keraal dataset is a medical database of clinical patients carrying out low back-pain rehabilitation exercises \cite{Nguyen2024IJCNN}. The data includes recordings from healthy subjects but, more importantly, of rehabilitation patients, extracted from a 4 weeks evolution of each patient.   
The centrally randomized, controlled, single-blind, and bi-centric study was conducted from October 2017 to May 2019.
The rehabilitation program includes a group of 31 patients, aged 18 to 70 years, recruited in the double-blind study. 
12 patients {suffering from low-back pain} were included in the Robot Supervised Rehabilitation Group, and were asked by a humanoid robot coach to perform each of the three predefined exercises the best they can from its demonstration.
 The details of this clinical trial, including the patient care, the rehabilitation sessions, the robot coach, the inclusion and exclusion criteria, the characteristics of the patients, and the efficiency of the care have been reported in \cite{Blanchard2022BRI}.
Details can be read on \url{http://nguyensmai.free.fr/KeraalDataset.html}.
A list of three exercises has been chosen in conjunction with therapists as common rehabilitation exercises that are also used for low-back pain treatment. 

Videos collected from patients and healthy subjects were annotated by two physiotherapists.  In this study, the labels are obtained by merging the assessments of two physicians, and we process the videos to obtain the BlazePose and OpenPose skeletons. 
Each exercise was labeled as either correct or incorrect. The dataset used in this study comprises Kinect (3D) v2 skeleton data , Blazepose (3D) and OpenPose (2D) skeletons obtained from videos and annotations.

\subsection{Kimore dataset}

The Kimore dataset \cite{capecci2019kimore} includes RGB-D videos and score annotations of five exercises for LBP rehabilitation, selected by physicians. The exercises are performed by two groups of participants: a control group (44 participants) and a group of patients (34 participants). The dataset also contains an assessment of the performed exercises, provided by two physicians.
More details can be found here \url{https://vrai.dii.univpm.it/content/kimore-dataset}.

\section{METHODOLOGY}
\label{sec:method}
This section provides the technical overview of two algorithms used in this study.

\subsection{Gaussian Mixture Model}
Gaussian mixture models (GMMs) belong to a group of probabilistic models used to classify data into different categories based on probability distribution. GMM models the dataset as a mixture of several Gaussian distributions.
As in \cite{Nguyen2016IWCR}, we encode the movement point positions as a Gaussian Mixture Model (GMM): $\theta= [t,x]$, where $t$ is the timestamp and $x$ the joints positions.

\begin{equation}
p({\theta}) = \sum_{i=1}^K\phi_i \mathcal{N}({\mu_i,\Sigma_i})
\end{equation}
where the $i^{th}$ vector component is characterized by normal distributions with weights $\phi_i$, means ${\mu_i}$, and covariance matrices ${\Sigma_i}$. 
Each Gaussian of the mixture is thus defined by: 
\begin{equation}
\mu_i = \left[
\begin{array}{c}
\mu_i^t\\
\mu_i^x
\end{array}
\right] 
,
\Sigma_i = \left[
\begin{array}{c c}
\Sigma_i^t & \Sigma_i^{xt} \\
\Sigma_i^{xt} & \Sigma_i^x
\end{array}
\right]
\end{equation}
where the indices $t$ and $x$ refer to respectively time and position.

The parameters $\phi_i, \mu_i, \Sigma_i$ are learned by Expectation-Maximisation (EM) from the skeleton data of the movements captured by the Kinect or estimated with OpenPose or BlazePose.

\subsection{Graph Convolutional Networks}

Graph Neural Networks (GNNs) are a class of deep learning models that are specifically designed to operate on graph-structured data \cite{Du2021AssessingPR}. These models leverage the graph topology to learn meaningful representations of the nodes and edges of the graph. 

Given our focus on skeletons, let us examine how they can be integrated into graph data. Skeleton-based data can be obtained from motion-capture devices or pose estimation algorithms from videos. Usually, the data is a sequence of frames, each frame will have a set of joint coordinates. Each joint in the given skeleton can be represented as a node in a graph, while connections between the joints represent the edges in the graph (e.g., right hip to right knee). In this way, the graph provides information about the hierarchy of the human skeleton, starting from one joint as a root (e.g., mSpine) and expanding further to hands and feet, which would be the leaves of the graph. Although GNNs have been extensively used in various domains, they were first used on static graph data, where graph structure does not change once data is fitted. In recent years, there has been an increased interest in systems with temporal dimension, meaning graph data would change over time. To address this need, a new family of GNNs has emerged: Spatio-Temporal GNNs, which take into account both the spatial and temporal dimensions of the data by learning temporal representations of the graph structure.

\begin{figure}
    \centering
    \includegraphics[width=\columnwidth]{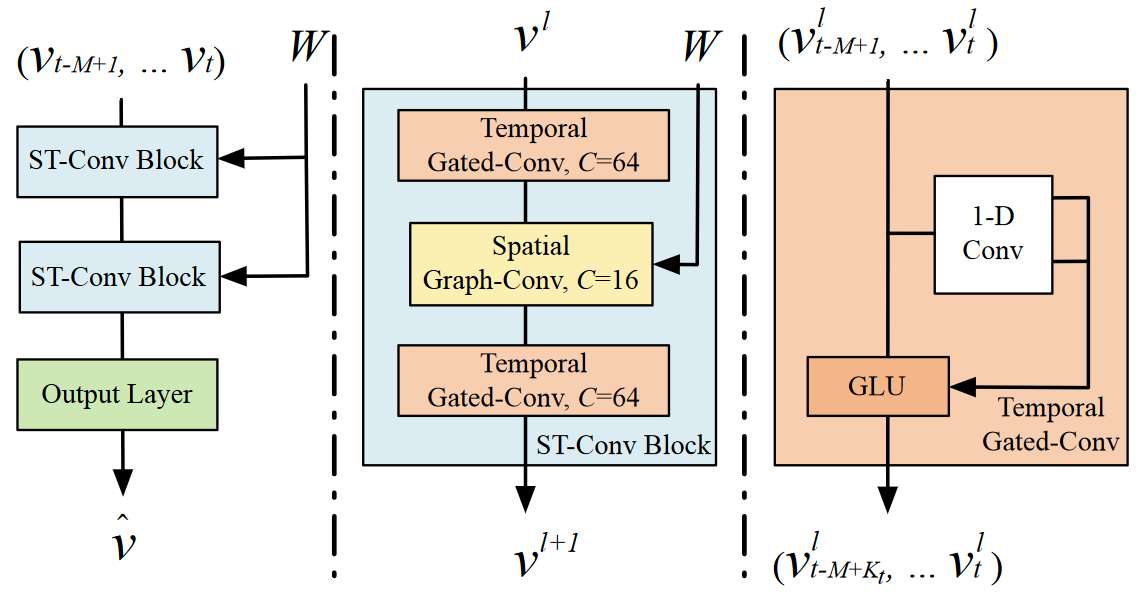}
    \caption{Architecture of spatio-temporal graph convolutional networks. The network chains two spatio-temporal convolutional blocks (ST-Conv blocks) and a fully-connected output layer. Each ST-Conv block contains two temporal gated convolution layers and one spatial graph convolution layer in the middle. Image taken from \cite{yu2018spatio}.}
    \label{fig:stgcn}
\end{figure}

Such architecture was first introduced in \cite{yu2018spatio}. The authors proposed the architecture of spatio-temporal graph convolutional networks (STGCN). As shown in Figure \ref{fig:stgcn}, STGCN is composed of several spatio-temporal convolutional blocks, each of which is formed as a “sandwich” structure with two gated sequential convolution layers and one spatial graph convolution layer in between.

\begin{figure}
    \centering
    \includegraphics[width=\columnwidth]{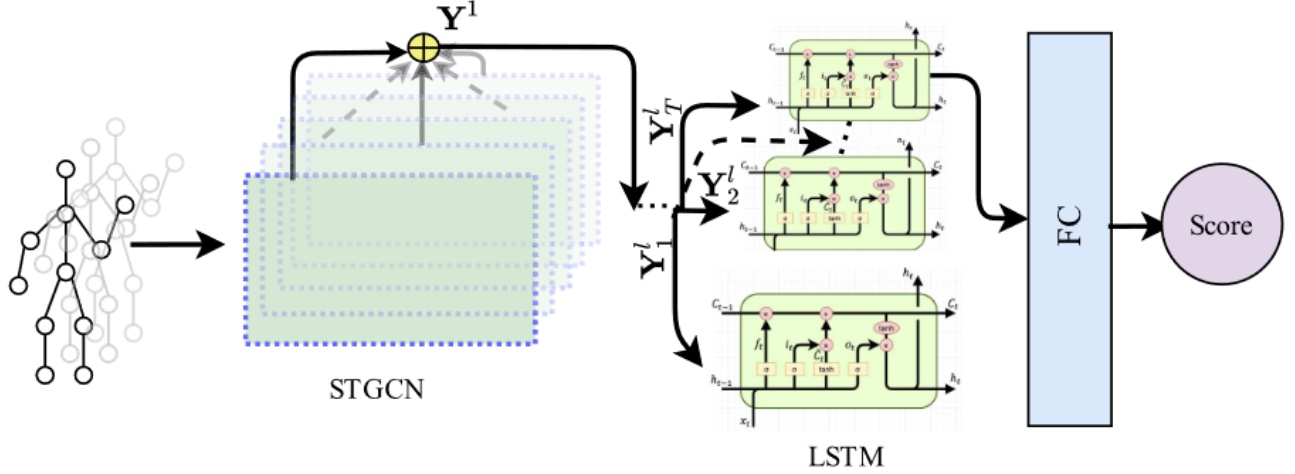}
    \caption{Architecture of STGCN for physical rehabilitation assessment. The network combines STGCNs with LSTM layers as suggested in \cite{deb2022graph}. Image taken from \cite{deb2022graph}.}
    \label{fig:stgcn_pr}
\end{figure}

Further, in \cite{yan2018spatial} this model was applied to skeleton-based action recognition, while \cite{deb2022graph} modified that algorithm for the task of assessing physical rehabilitation exercises. Since it is a very novel approach obtaining state-of-the-art results, we decided to their algorithm as the base for our analysis here. An overview of this model can be seen in Figure \ref{fig:stgcn_pr}. 

\section{RESULTS}
\label{sec:results}
\begin{figure*}
     \centering
     \begin{subfigure}[b]{0.3\linewidth}
         \centering
         \includegraphics[width=\textwidth]{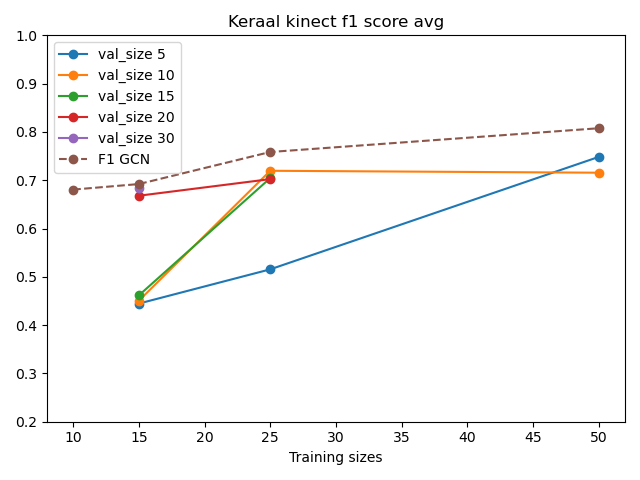}
         \caption{F1 scores with Kinect v2}
         \label{fig:kinect_gmm_keraal}
     \end{subfigure}
     \hfill
     \begin{subfigure}[b]{0.3\linewidth}
         \centering
         \includegraphics[width=\textwidth]{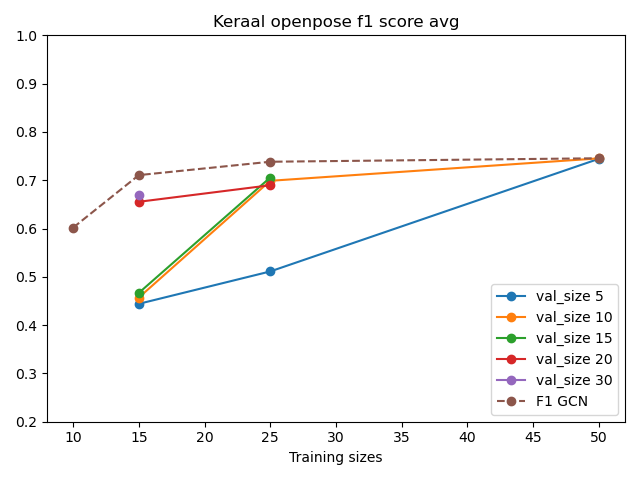}
         \caption{F1 scores with OpenPose}
         \label{fig:openpose_gmm_keraal}
     \end{subfigure}
     \hfill
     \begin{subfigure}[b]{0.3\linewidth}
         \centering
         \includegraphics[width=\textwidth]{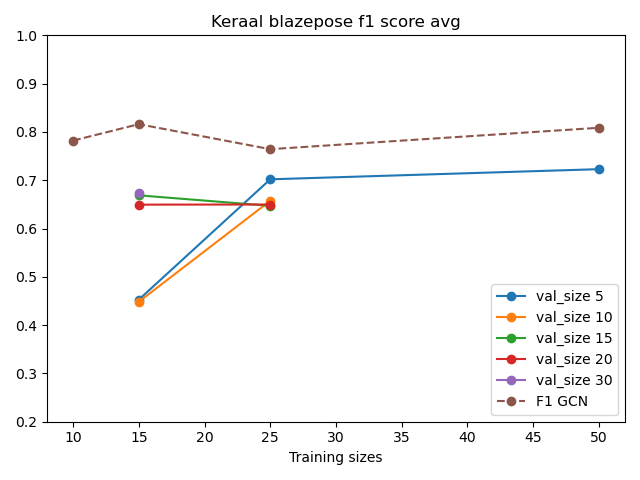}
         \caption{F1 scores with BlazePose}
         \label{fig:blazepose_gmm_keraal}
     \end{subfigure}
  \vspace{-0.4em}
        \caption{F1 scores of GCN and GMM on the Keraal dataset, with different training sizes used. The data is averaged across exercises or groups. For GMM various sizes of validation sizes (for threshold defining) are deployed.
        }
        \label{fig:gmm_keraal}
\end{figure*}

\begin{figure*}
     \centering
     \begin{subfigure}[b]{0.3\linewidth}
         \centering
         \includegraphics[width=\textwidth]{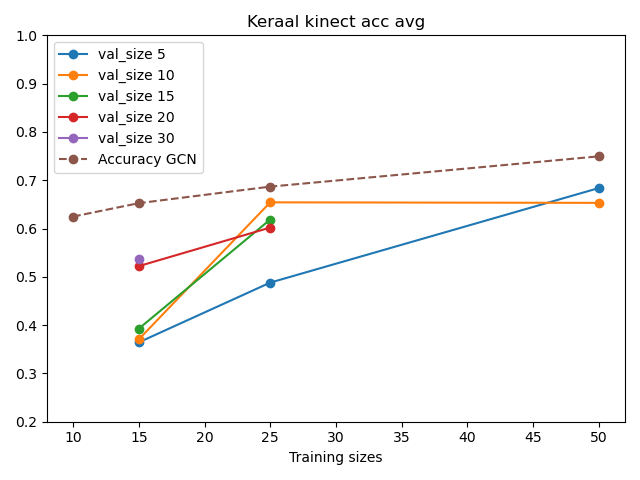}
         \caption{Kinect v2}
         \label{fig:kinect_gcn_keraal}
     \end{subfigure}
     \hfill
     \begin{subfigure}[b]{0.3\linewidth}
         \centering
         \includegraphics[width=\textwidth]{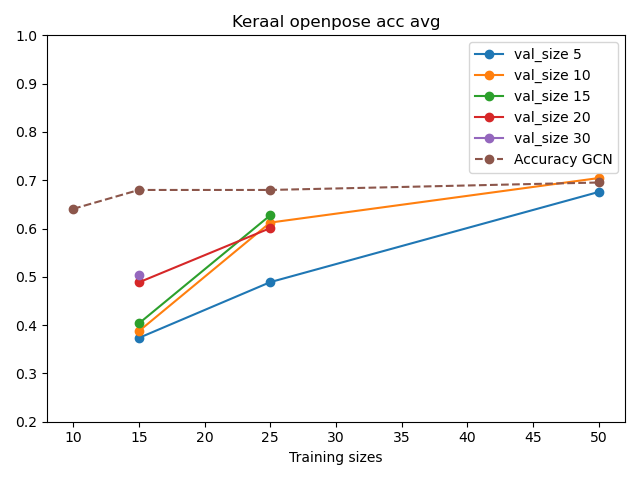}
         \caption{OpenPose}
         \label{fig:openpose_gcn_keraal}
     \end{subfigure}
     \hfill
     \begin{subfigure}[b]{0.3\linewidth}
         \centering
         \includegraphics[width=\textwidth]{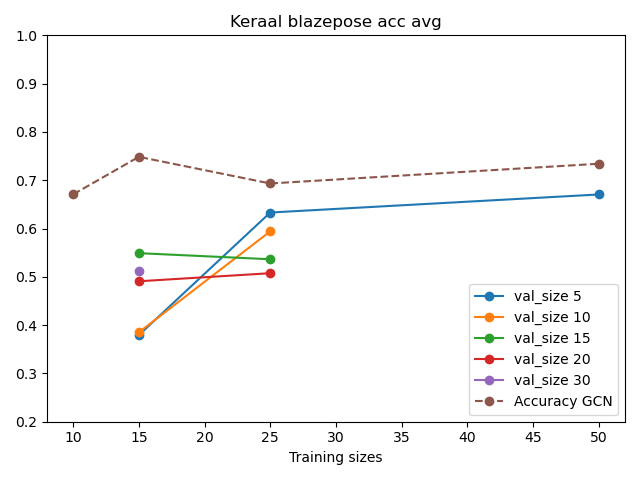}
         \caption{BlazePose}
         \label{fig:blazepose_gcn_keraal}
     \end{subfigure}
  \vspace{-0.4em}
        \caption{Accuracy of STGCN and GMM on the Keraal dataset.
        }
        \label{fig:gcn_keraal}
\end{figure*}

We experimented with the STGCN and GMM algorithms and compared their performances and sample efficiency on the two physical rehabilitation exercises datasets.
The GMM is trained on correct demonstrations of the exercises, and then a classification threshold is determined based on validation data containing both correct and incorrect demonstrations. In contrast, the STGCN method needs to be trained on both correct and incorrect demonstrations. Even though validation data could be used to optimize hyperparameters or to perform early stopping, we did not use it with this method. To compare the data efficiency of both algorithms, we thus need to take into account both the number of training and validation examples needed for the GMM method and compare it with the number of training examples needed for the STGCN based method. We report the scores after training, which takes a couple of seconds for GMM to train, while STGCN, for one training of 250 epochs, takes 20 - 70 minutes depending on the setup (which dataset, skeleton type, and number of training examples). All models have been trained on CPU Intel Core i9-9900KF.

Figure \ref{fig:gmm_keraal} shows the F1 scores obtained with two methods on the Keraal dataset, while \ref{fig:gcn_keraal} provides the F1 scores. The scores are averaged across the 3  exercises of the dataset, with Kinect, OpenPose, and BlazePose poses. We can notice a slight but not significant improvement in these scores as the training set size increases. For both GMM and STGCN, we note that the scores with Kinect, Openpose and BlazePose are similar : the use of depth sensors (Kinect v2) does not seem to improve significantly the performance of the algorithm, which corroborates the conclusions presented in \cite{marusic2023evaluating}:  for low-back rehabilitation exercises, previous GMM obtained through Kinect, OpenPose and BlazePose data revealed comparable results. These new results extend the same conclusion to variations in the sizes of the training and validation sets, and to another evaluation algorithm : STGCN. This indicates that independently of the size of the dataset and the machine learning algorithm, simple RGB cameras have the potential to be used as the main sensor for collecting movement data.  
On Kinect and BlazePose data, the STGCN method seems to outperform the GMM method, especially when a large number of training examples are available. These results advocate in favor of using the STGCN method, even when few training examples are available.

\begin{figure}
    \centering
    \includegraphics[width=\columnwidth]{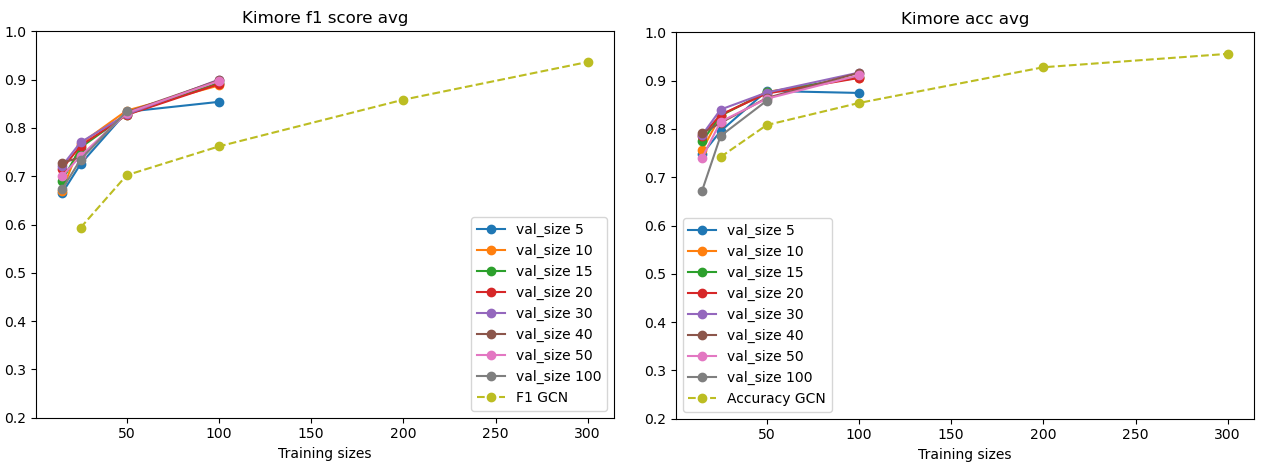}
    \caption{F1 score and accuracy scores of STGCN and GMM on the Kimore dataset.}
    \label{fig:stgcn}
\end{figure}


Results for the Kimore dataset are presented in Figure \ref{fig:stgcn}.  While GMM can achieve an F1 score of nearly 0.9 with 100 training examples, we can see that more than 200 examples are needed to achieve such classification precision with STGCN. Even when taking into consideration the additional validation examples needed for GMM, this model seems to perform significantly better when a small number of examples is available.
%
This result contradicts what was observed for the Keraal dataset, where the STGCN method outperforms the GMM method even with a few training examples. Our supposition is that this could be due to the absence of a strong agreement between the physicians for Keraal (Cohen’s $\kappa = 0.63$ and Krippendorff’s $\alpha = 0.62$). In comparison, the Kimore dataset uses a questionnaire containing 10 questions to assess the quality of the performed exercises, which could lead to more robust labeling.

\section{CONCLUSIONS}
\label{sec:conclusion}

In this work, we have compared two algorithms for LBP physical rehabilitation assessment on two datasets with several human pose estimation methods. We can draw several conclusions from the observed results, that can provide useful insights in order to further develop the use of automated physical rehabilitation methods :

\begin{itemize}
    \item While more experiments could be done in order to confirm this result, we observed that using more expensive depth cameras does not seem to impact the performance of the assessment method. This study confirms the conclusion presented in \cite{marusic2023evaluating} over a more extensive study using more sizes of training and validation sets and, additionally, using a more efficient evaluation algorithm. Similarly to that evaluation, we see that the use of 3D inputs (Kinect and BlazePose), compared to 2D inputs (OpenPose) does not improve the results obtained on the Keraal dataset.

    \item More training examples lead to a better assessment. We recommend collecting data from as many participants as possible when recording exercises.
    \item Label quality is essential. We observe significantly better accuracy on the Kimore dataset, where the labels were obtained by merging the answers to ten questions given by two physicians, compared to the Keraal dataset, where only two evaluations are combined. Although this reveals that assessing rehabilitation movements is a difficult task, we suggest having as many annotators as possible and monitoring a measure of their agreement to ensure high label quality.
    \item Finally, we recommend using the STGCN algorithm instead of the GMM algorithm in most situations. The GMM algorithm should still be useful in special cases when we need a fast (real-time) learning system or when gathering incorrectly performed exercises is difficult. It can be trained using only correct demonstrations and only needs a few incorrect demonstrations to optimize the threshold value used for classification.
\end{itemize}









\bibliographystyle{plain}
\bibliography{sample-base}

\end{document}